# High-efficiency and broadband coherent optical comb generation in integrated X-cut lithium niobate microresonators

Yunxiang Song[1,2], Zongda Li[3,4], Xinrui Zhu[1], Norman Lippok[1], Miro Erkintalo[3,4], Marko Lončar[1,*]

[1] John A. Paulson School of Engineering and Applied Sciences, Harvard University, Cambridge, MA 02138, USA
[2] Quantum Science and Engineering, Harvard University, Cambridge, MA 02139, USA
[3] Department of Physics, University of Auckland, Auckland 1010, New Zealand
[4] The Dodd-Walls Centre for Photonic and Quantum Technologies, Dunedin, New Zealand
[*] Corresponding author: loncar@g.harvard.edu

**The ability to generate efficient and coherent frequency combs using photonic integrated circuits offers tremendous potential for a range of applications. In particular, "microcombs" based on chip-integrated resonators are poised to revolutionize optical communication, computation, and sensing systems, especially when paired with fast electro-optic (EO) devices. X-cut thin-film lithium niobate (TFLN) is a promising platform for developing the next-generation of microcomb-driven integrated photonic systems, providing a diversity of functionalities through combined $\chi^{(3)}$ and EO nonlinearities. In this context, normal-dispersion Kerr microcombs are critically needed because of their standout advantages, yet this dispersion regime remains unexplored for comb generation on X-cut TFLN. Here, we leverage ultralow-loss photonic waveguides, as well as strategic resonator designs that allow us to tailor Raman effects and engineer desired spatial mode interactions, for the robust generation of normal-dispersion Kerr microcombs. Specifically, using a microresonator with ~100 GHz free spectral range (FSR), we show Kerr microcombs that substantially surpass state-of-the-art bright cavity soliton and EO microcombs on X-cut TFLN in key performance metrics. Our microcombs exhibit <2 dB power variation across the telecom C-L bands, >50% pump-to-comb conversion efficiency, and <10 mW on-chip turn-on power. Additionally, we demonstrate a novel microcomb whose existence is underpinned by both normal-dispersion Kerr dynamics and stimulated Raman scattering, in a single spatial mode of a microresonator with ~25 GHz FSR. This microcomb manifests itself as two interleaved frequency combs centered about the pump and Stokes frequencies, resulting in extended spectra spanning nearly 33 THz (300 nm) in bandwidth. Our work will unlock high-speed and low energy consumption photonic circuits for communications, frequency synthesis, and signal processing enabled by a monolithic microcomb technology, while also stimulating further investigations of new nonlinear states that may synergize the strong hybrid nonlinearities unique to X-cut TFLN.**

# Introduction

Microresonator frequency combs are revolutionizing on-chip photonic systems[1–5]. Also known as "microcombs", these light sources enable wavelength-division multiplexed (WDM) applications such as communications[6–11], signal processing[12–15], and imaging[16,17], by offering a phase-locked array of equally spaced optical frequency lines. Microcombs generated under conditions of normal group-velocity dispersion are particularly well-suited for these applications by virtue of their unique characteristics: relatively flat spectral profile, high pump-to-comb conversion efficiency, and ease of generation[18–24]. On the other hand, the manipulation of comb lines required by many such applications critically rely on high-performance electro-optic (EO) components. Because of this, microcomb-driven integrated photonic systems to-date typically involve multiple photonic chips: one dedicated to comb generation, and others, often silicon photonics-based, for comb processing[4,9,10]. Such heterogeneous integration has certain advantages, for example the comb generation chip can utilize highly nonlinear and ultralow-loss materials, whereas the silicon chips benefit from technologically mature, high-volume, and cost-effective manufacturing. However, it also comes with its challenges, including increased energy consumption due to substantial insertion losses, and larger system complexity caused by mismatched material platforms. The absence of a monolithic solution that may overcome these challenges stems from fundamental limitations: coherent microcombs at telecommunications-, biological-, and quantum-relevant wavelength bands cannot be generated on silicon, while the leading materials for microcombs lack efficient EO modulation capabilities[24–31].

In search for a unifying material, X-cut thin-film lithium niobate (TFLN) and thin-film lithium tantalate (TFLT) have recently emerged as promising platforms supporting both high-speed and low energy consumption EO at the chip-level[32–36]. The former platform is particularly interesting, as it is already commercially manufactured at scale by semiconductor foundries. The high repeatability and scalability of X-cut TFLN photonic circuit fabrication has potential to offer large-scale photonic systems driven by monolithically integrated coherent microcombs. Despite the continued pursuit for such a comb source, however, resonant EO combs on X-cut TFLN (and TFLT) face numerous drawbacks, such as exponential optical power roll-off with increasing mode number, prohibitively large electrical power consumption, and sensitivity to resonance matching requirements[37–40]. Although the strong Kerr nonlinearity of these materials has recently been used to realize anomalous-dispersion soliton microcombs[34,41–43], bright solitons are still limited by their non-flat spectral shapes, low pump-to-comb conversion efficiency, and often the need for bulky auxiliary equipment to sustain comb generation. While normal-dispersion Kerr microcombs could alleviate all these shortcomings, they have yet to be realized on X-cut TFLN due to the platform's strong and complex Raman response[43–45], whose precise effects on normal-dispersion Kerr dynamics remain largely unexplored.

In this work, we overcome the challenges outlined above and demonstrate normal-dispersion Kerr microcombs on X-cut TFLN. Our results are enabled by judicious balancing of the host microresonator's ordinary ($n_o$) to extraordinary ($n_e$) index ratio, which allows us to tailor the strength of the Raman effect (relative to the Kerr effect) based on the material's anisotropic Raman response[42,43]. Equally important, we integrate non-adiabatic waveguide mode converters into our microresonators, so as to controllably initiate normal-dispersion Kerr dynamics in the continuous-wave (CW) driven soft excitation regime. Leveraging these advances, we achieve record-high quality-factors for nonlinear microresonators on X-cut TFLN, which underscores the characteristic high-performance of normal-dispersion microcombs. Specifically, we demonstrate a 102.4 GHz repetition-rate X-cut TFLN normal-dispersion Kerr microcomb, with a spectral bandwidth over 24 THz (200 nm), measured pump-to-comb conversion efficiency exceeding 50%, and an on-chip turn-on power as low as ~8 mW. We conduct measurements of the optical linewidth of the normal-dispersion Kerr combs and find quiet modes[46] which feature substantial reduction in Lorentzian linewidth compared to that of the pump laser. This result hints to a unified description for pump noise transduction into the comb line fluctuations, underlying Kerr combs in both anomalous-[46–48] and normal-dispersion regimes. We also identify a new kind of normal-dispersion microcomb, arising via a synergetic interplay between the Kerr and stimulated Raman scattering (SRS) effects, which corresponds to two interleaved combs spectrally reminiscent but mechanically distinct from the so-called Stokes solitons in the anomalous dispersion regime[49] and bi-chromatically pumped cavity solitons[50,51]. This finding contrasts with the dominant perspective that strong, narrowband SRS hinders coherent microcombs on crystalline photonic materials[52]. Our experimental observations are corroborated and informed by numerical simulations of the generalized Lugiato-Lefever equation (gLLE), tuned to the salient experimental parameters. The excellent agreement between experiment and simulation suggests that SRS, combined with the normal-dispersion Kerr effect, can support novel nonlinear states that are characterized by very broad spectral profiles and simultaneously microwave-level comb spacings.

## Results

### *Device design*

Integrated X-cut TFLN microresonators are known to feature strong SRS, meaning resonance modes that are Stokes-shifted from the pumped modes experience large effective gain. In the CW-pumped regime, intracavity SRS gives rise to efficient lasing action[44,53], suppressing modulation instability (MI) gain responsible for Kerr microcombs. However, previous Raman spectroscopy measurements reveal that the Raman response is anisotropic, attaining its minimum (maximum) value when light is polarized along the $n_o$ ($n_e$) axis of X-cut TFLN's crystal plane[42,43]. Thus, to suppress parasitic SRS and achieve high-performance normal-dispersion Kerr microcombs, we leverage this fact and design microresonators in racetrack configuration. Such designs can strongly bias light propagation in one crystal axis over the other, thus minimizing the round-trip Raman response (relative to the Kerr effect, as quantified by the Raman fraction $f_R$ in gLLE simulations) [Fig. 1(a)]. According to this, we fabricated two sets of racetrack microresonators featuring different free spectral ranges (FSRs) of 102.4 GHz and 25.7 GHz, that are further described below.

A key component for initiating normal-dispersion Kerr dynamics under CW-driving is local resonance frequency perturbations as seen by the pump, often realizable by engineering avoided mode crossings (AMXs) into the microresonator's dispersion profile[54–60]. Such AMXs can be achieved via spatial mode interactions[19,54,55], coupled resonator configurations[21–23], or photonic crystal designs[24]. While common wisdom suggests that the disadvantage of the spatial mode interactions approach is the difficulty to systematically achieve and control the AMX size, we overcome this challenge by introducing a rapidly tapering waveguide segment within the microresonator, i.e., connecting two highly mismatched widths over a short distance [illustrated in Fig. 1(b)]. This design scatters a fraction of the fundamental transverse-electric-like ($TE_0$) waveguide mode into the higher-order $TE_1$ mode by virtue of its non-adiabaticity, in a manner that is both systematic and controllable, i.e., repeatable from device to device. To verify this, we measured the electric field coupling coefficient that characterizes the relative resonance splitting at an AMX, for eight different microresonators (four 102.4 GHz-FSR and 25.7 GHz-FSR devices each), obtaining consistent values across each measurement [Fig. 1(c)]. Since this coupling is constant regardless of device FSR, our measurements confirm that the AMX is created by the non-adiabatic tapered waveguide alone.

Figures 1(d-f) and 1(g-i) show illustrative examples of the measured dispersion and intrinsic quality factors ($Q_i$) of the microresonators used in subsequent experiments. In Fig. 1(d) and 1(g), we specifically plot the integrated dispersion defined as $D_{int}(\omega_\mu) = \omega_\mu - \omega_0 - \mu \cdot D_1 = \sum_{n\geq 2} \frac{D_n}{n!} \mu^n$, which describes the deviation of cavity resonance frequencies $\omega_\mu$ from an equidistant grid of spacing $D_1 = 2\pi \cdot f_{FSR}$ about center frequency $\omega_0$. As can be seen, both curves exhibit globally a parabolic shape ($D_2 < 0$), which indicates normal group-velocity dispersion. However, strong dispersion perturbations due to the AMXs are also clearly visible, noting especially the frequency red-shifted branch of modes near ~1547 nm for the 102.4 GHz-FSR device and near ~1592 nm for the 25.7 GHz-FSR device. Such AMXs and their associated resonance splittings are highlighted in the higher order dispersion curves [Figs. 1(e) and 1(h)], where the dominant $D_2$ term is removed from $D_{int}$.

Despite the non-adiabatic tapered waveguides that supply the AMXs, we find that excess losses associated with them are very low. The average $Q_i$s for the 102.4 GHz-FSR and 25.7 GHz-FSR devices are 5.16 million and 7.57 million considering resonance modes in the telecom C-L band, respectively, with highest values reaching nearly 15 million in both devices [Figs. 1(f) and 1(i)]. Compared to other nonlinear experiments that use X-cut TFLN, our Q-factors are extraordinarily high, which is primarily due to the relatively wide waveguide top widths permitted by operation in the normal dispersion regime. As will be seen, these high-Q factors further accentuate the characteristically high performance associated with normal dispersion Kerr microcombs, especially towards low-threshold microcomb generation.

*Normal-dispersion Kerr microcomb*

We may expect that the AMXs present in our devices' dispersion profiles can facilitate the formation of microcombs. To validate this position, we first consider the 102.4 GHz-FSR microresonator within our characterization setup [Fig. 2] and optically drive it with a CW laser, whose power is set to ~8 mW on-chip and wavelength is set to coincide with the second-most red-shifted mode of the ~1547 nm AMX. By sweeping the pump laser around this resonance mode, we observe that Kerr microcombs emerge. The process is clearly manifested in an abrupt step-change of the transmitted power contained in the comb and pump frequencies, and is bidirectional, occurring regardless of whether the laser is swept from blue to red or vice versa [Figs. 3(a)]. Notably, when initiating the laser scan from red detuning, intermediate nonlinear states are absent and the broadest possible microcomb spontaneously forms.

The blue curve in Fig. 3(b) shows the measured optical spectrum when the detuning between the pump laser and the resonance mode is maintained at a fixed position within the step region marked in Fig. 3(a). Also shown as the red curve is the spectrum obtained from numerical simulations of the gLLE that use experimental parameters. The simulations and experiments are in excellent agreement, both featuring spectral asymmetry that is primarily due to the set of AMX-shifted modes near the pump, which modifies the phase matching condition. Nevertheless, the comb exhibits high spectral flatness across the telecom C- and L-bands, containing 19 and 14 comb lines with less than 2 dB power variation in each band. In the time domain, our simulations predict that the comb structure corresponds to an ultrashort pulse that sits atop a broad pedestal [Fig. 3(b) inset]. In stark contrast with normal-dispersion "dark solitons" as conventionally defined[61–64], we find that the upper and lower power levels of the pulse do not correspond to the CW steady-states of the gLLE, nor is the pulse's existence confined to the bistable parameter regime, as evidenced by its spontaneous emergence in the red to blue detuning scan.

We verify the mutual coherence between the comb lines by electro-optically dividing down the 102.4509 GHz repetition-rate beatnote to an electronically detectable microwave signal [Fig. 3(c)]. The strong microwave beatnote has no spurious frequency components, indicative of one comb line per resonance mode and high mutual coherence between the comb lines.

To gain more insight into the coherence properties of comb, we measure the optical linewidth of selected comb lines using delayed self-heterodyne interferometry (see Supplementary note for details). The Lorentzian (intrinsic) linewidths (denoted $\Delta \nu$) can be estimated from the measured frequency noise power spectral density (PSD) of each comb line [Fig. 3(d) shows four example PSDs], by directly taking the white frequency noise floor[46]. Values at lower frequency offsets are dominated by technical contributions[46,48] such as thermorefractive noise, that can shift resonance positions at slow time-scales, thus they do not reflect the fundamental noise scaling with comb line number for a fixed detuning state. As shown in Fig. 3(e), we find that the $\Delta \nu$ features a parabolic distribution, with the minimum of the parabola centered about mode -11 and positioned far below the $\Delta \nu$ of the pump laser (consistent with $\Delta \nu$ of mode 0 of the comb). This behavior is largely in agreement with previous characterizations of CW-driven dissipative solitons in integrated $Si_3N_4$ microresonators, except that here, modes with suppressed noise are blue-shifted from the pump[46]. Our results point to a rather universal mechanism of noise scaling, where frequency fluctuations of the pump laser reflect onto the comb lines via the influence of effective detuning on comb repetition-rate, in our case determined by the collective dispersive wave recoil arising from numerous AMXs.

In addition to the excellent comb coherence, normal-dispersion Kerr microcombs are also known for their high pump-to-comb conversion efficiency. By tuning the pump laser across the comb's existence range, we change the effective detuning governing the comb state and collect distinct spectra [Fig. 3(f) inset]. We find that as the effective detuning decreases, the spectrum broadens, and the comb power increases, accompanied by pump power extinction and increased pump-to-comb conversion efficiency [Fig. 3(f)]. For the least detuned state, we measured conversion efficiencies as high as 54%, thanks to operating in both the blue-detuned and twice overcoupled regimes. Compounded by a turn-on power less than 10 mW, our normal-dispersion Kerr microcombs surpass several key performance metrics of state-of-the-art microcombs on X-cut TFLN (see Supplementary note for detailed comparison).

*Normal-dispersion Kerr and Raman Stokes microcombs*

While the results described above pertain to the devices with comparatively large, 102.4 GHz FSR, we find that similar normal-dispersion Kerr microcombs (and associated comb generation dynamics) also manifest themselves in microresonators with 25.7 GHz FSR. Indeed, Fig. 4(a) shows exemplary spectra of such comb states, including 1-FSR spacing combs and crystal-like states with comb line spacing an integer multiple of the FSR (Fig. 4(a) right panel shows an example of a crystal state with 5-FSR spacing). These combs are obtained when optically driving the second-most red-shifted mode of the device's 1592 nm AMX, and with 78 mW of CW-power on-chip. Their spectral envelopes show characteristic sinc-like spectral lobes around the pump wavelength, suggesting quasi-rectangular temporal pulse shapes. Notably, finite higher order dispersions contribute to additional asymmetry in the spectral wing positions about the central spectral lobes.

Interestingly, while Raman effects were found not to play a role for the operating conditions of our 102.4 GHz-FSR device, our experiments reveal that they may drive the emergence of novel dynamical states in our 25.7 GHz-FSR device. The marked difference in their respective state spaces lies mainly in their different Raman spectra, which is sensitive to the microresonators' $n_o$ to $n_e$ ratio (see Supplementary note for an expanded discussion). In the 25.7 GHz-FSR case, the primary Kerr microcomb acts as a multi-tone source, thereby supplying a combined Raman gain to Stokes-shifted modes outside of the normal-dispersion comb's original bandwidth. Above threshold, additional frequency components can be generated, giving rise to two qualitatively distinct regimes of microcomb operation which will be elaborated upon below.

Figure 4(b) shows an illustrative optical spectrum in which a normal-dispersion Kerr microcomb around the pump wavelength of ~1592 nm coexists with a strong Stokes component centered at ~1762 nm. Also depicted (red curve) is the corresponding result from numerical simulations, showing good agreement with experiments. Similar to chaotic MI combs in the anomalous-dispersion regime, we find that electronic detection of the comb in Fig. 4(b) yields a substantially diminished beatnote and an elevated noise floor [Fig. 4(c)]. This observation suggests that the comb is unstable, exhibiting amplitude and phase fluctuations over time. Our numerical modelling corroborates the chaotic nature of the comb: in the time domain, the state corresponds to a picosecond-scale bright pulse with complex internal structure that fluctuates from round-trip-to-round-trip [Figs. 4(d) and (e)].

Strikingly, we find that as the pulse becomes shorter in time (as evidenced by the comparatively wider spectral lobes around the ~1592 nm pump), the chaotic state can transition into an alternate, stable structure. While the spectral envelope of this state [Fig. 4(f)] resembles that shown in Fig. 4(b), the state is clearly distinguished by the clean and strong electronic beatnote and a noise floor at the measurement limit [Fig. 4(g)]. These electronic measurements indicate high coherence of the comb lines.

The stable state demonstrated in Figs. 4(f) and 4(g) can also be observed in our simulations [red curve in Fig. 4(f)]. The simulations [Figs. 4(h) and 4(i)] show that the corresponding temporal waveform consists of a picosecond-long quasi-rectangular pulse, carrying rapid oscillations on its otherwise flat top with a period of about $1/\Omega_R$, where $\Omega_R \sim 18.8$ THz is the Stokes shift. These oscillations arise from the beating between the spectral features around the pump and their Stokes-shifted Raman components. Moreover, the oscillations exhibit a temporal drift relative to the background pulse envelope, signaling a phase-velocity mismatch between the center frequencies of the primary Kerr comb and the Raman Stokes comb. This in turn indicates that, while the two combs share the same line spacing, their carrier envelope offset frequencies are different. In other words, the two combs are interleaved, spectrally reminiscent of (but mechanically distinct from) "Stokes solitons" observed in anomalous-dispersion silica resonators[49] and spectrally-extended Kerr combs generated in various systems with more than one optical pump[50,51].

## Discussion

In summary, we have demonstrated normal-dispersion Kerr microcombs on X-cut TFLN for the first time, leveraging racetrack designs and embedded non-adiabatically tapered waveguides which suppress parasitic SRS and supply controllable AMX, respectively. Thanks to operation in the normal dispersion regime, where devices with ultrahigh quality-factors are feasible, our combs exhibit low threshold powers while also offering high pump-to-comb conversion characteristics. Furthermore, we have experimentally observed and numerically studied novel Raman-enhanced states, including coherent states comprised of two interleaved frequency combs and ultrabroad total bandwidths.

Our results represent a significant advance towards realizing high-efficiency on-chip optical frequency combs in an EO-compatible platform. The proven ability to suppress parasitic Raman effects will enable Kerr nonlinearity-based photonic devices, such as parametric oscillators[65,66] and optical isolators[67]. The realization of microcombs in the normal dispersion regime will allow protective cladding materials to be considered in the microresonator design, which in turn will facilitate fabrication via standardized X-cut TFLN foundry processes. Thus, the next-generation of integrated photonic transceiver and signal processing systems can utilize the efficient and low-noise WDM grid provided by the normal-dispersion Kerr microcomb, as well as compact, high-speed, and low energy consumption EO drivers[68,69], fabricated on a single chip with minimal interconnect loss. Besides a multitude of application benefits that the X-cut TFLN platform can offer in concert with our state-of-the-art comb source, we envisage that the novel Raman-mediated comb states observed in our experiments will attract substantial fundamental interest in the context of nonlinear dynamics and dissipative solitons. The strong Raman Stokes comb adds to the possible ways that frequency combs may be generated in exotic wavelength bands and may find use in metrology and spectroscopy. Inspired by this result, future microcomb generation and control methods on X-cut TFLN may look forward to fully harnessing its myriad nonlinearities, including SRS, Kerr, EO, $\chi^{(2)}$, and Brillouin effects[70–77].

**Acknowledgements**: We thank Prof. Federico Capasso, Prof. Stéphane Coen, Prof. Evelyn Hu, Prof. Stuart Murdoch, Prof. Victor Torres-Company, Dr. Guanhao Huang, Donald Witt, Dr. Yiqing Xu, and Yan Yu for helpful discussions. We thank Dr. Yuqi Zhao, Dr. Yu Zhang, and HyperLight Corporation for providing the X-cut TFLN EOM.

**Funding**: We acknowledge financial support from the Air Force Office of Scientific Research (FA955024PB004), the Department of the Air Force (FA8702-15-D-001), the National Science Foundation (ECCS-2407727, ECCS-2025158), the Department of Defense (FA9453-23-C-A039), and the Marsden Fund of the Royal Society Te Apārangi of New Zealand.

**Competing interests**: M. Lončar is involved in developing lithium niobate technologies at HyperLight Corporation. The authors declare no other competing interests.

**Data and materials availability**: All data needed to evaluate the conclusions in the paper are present the paper and/or the Supplementary note.

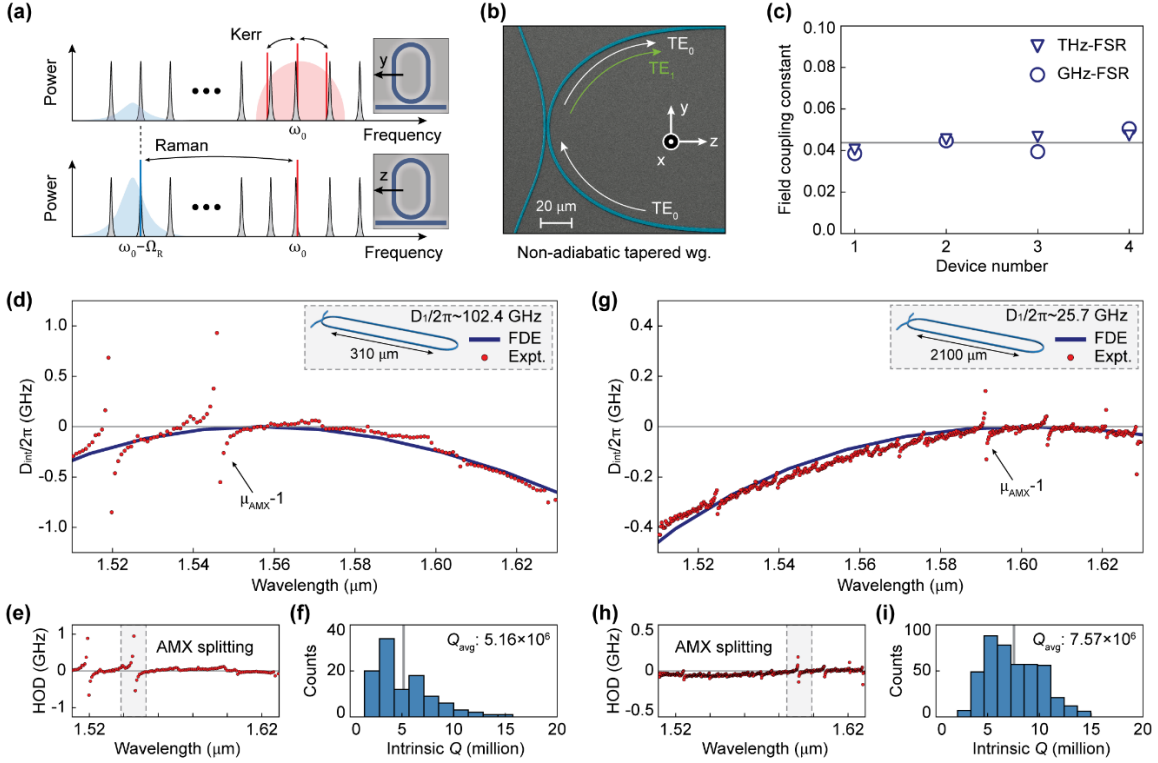

**Fig. 1 | Design and linear characterization of normal-dispersion microcomb sources. (a)** Nonlinear frequency generation in X-cut TFLN racetrack microresonators. Top panel illustrates modulation instability gain (red shaded region) and normal-dispersion comb formation around the pump $\omega_0$ (red line), resulting from the continuous-wave (CW) driving of a resonance mode near an avoided mode crossing (AMX). Suppressed Raman gain (blue shaded region) owing to the racetrack microresonator's orientation relative to the crystal axes (inset) is insufficient to overcome cavity losses. Bottom panel depicts strong Raman gain in an orthogonally oriented racetrack microresonator, and Raman lasing action near the Stokes-shifted mode $\omega_0 - \Omega_R$ (blue line) is induced by the CW drive. In both panels, gray triangles symbolically represent resonance modes. **(b)** Scanning electron microscope image of the non-adiabatic tapered waveguide, which supplies AMXs. Waveguide regions are highlighted in blue (false color). Specifically, it performs weak $TE_0$ to $TE_1$ mode conversion via a compact Euler bend design with minimum radius of 50 µm, connecting a universal 1.2 µm-wide bus-resonator coupling section to a 2.45 µm-wide (2.30 µm-wide) straight propagation section, of the 102.4 GHz-FSR (25.7 GHz-FSR) device. **(c)** Measured electric field coupling coefficient between the $TE_0$ and $TE_1$ modes in microresonators with 102.4 GHz and 25.7 GHz FSR, labeled by blue triangles and circles, respectively. **(d-i)** Integrated dispersion ($D_{int}$), higher order dispersion (HOD, or $D_{int} - \frac{D_2}{2}\mu^2$), and intrinsic quality factor statistics of the 102.4 GHz-FSR **(d, e, f)** and 25.7 GHz-FSR **(g, h, i)** devices used for microcomb generation. In (d) and (g), normal group-velocity dispersions are achieved using top widths as earlier mentioned in (b), combined with 0.337 and 0.330 µm waveguide heights and ~67 degrees sidewall angle, respectively. Further, the device FSRs ($D_1/2\pi$) are set by the straight propagation section lengths, being 310 µm and 2100 µm, respectively. FDE stands for finite-difference eigenmode simulation. Finally, the red-shifted pump modes considered in Figs. 3 and 4 are marked by the black arrows.

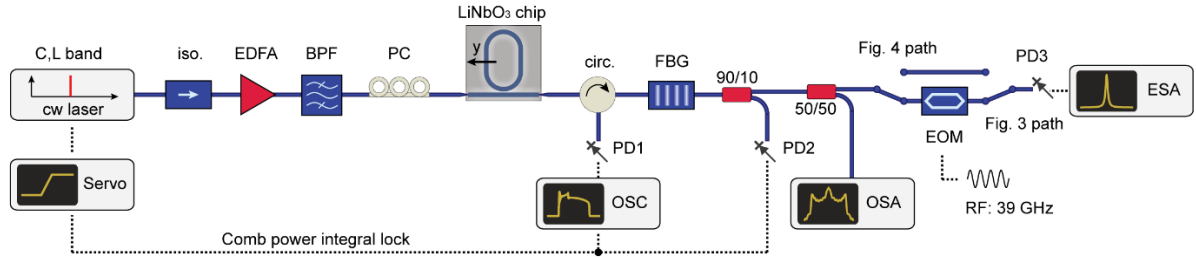

**Fig. 2 | Setup for microcomb generation.** Schematic illustration of the experimental setup. Abbreviations are listed as follows: iso (isolator); EDFA (erbium-doped fiber amplifier); BPF (bandpass filter); PC (polarization controller); circ (circulator); FBG (fiber Bragg grating); EOM (EO modulator); RF (radiofrequency); ESA (electrical spectrum analyzer); OSA (optical spectrum analyzer); OSC (oscilloscope); PD (photodetector). We note that PD1, PD2, and PD3 have electrical 3-dB bandwidths of 0.125, 0.125, and 43 GHz, respectively. The CW laser frequency can be actuated by an error signal corresponding to comb power fluctuations, with 50 Hz-integral feedback limit (Servo). Between the 50/50 power splitter and PD3, the bottom path labeled "Fig. 3 path" is used to characterize the 102.4 GHz-FSR device, and the top path labeled "Fig. 4 path" is used for the 25.7 GHz-FSR device.

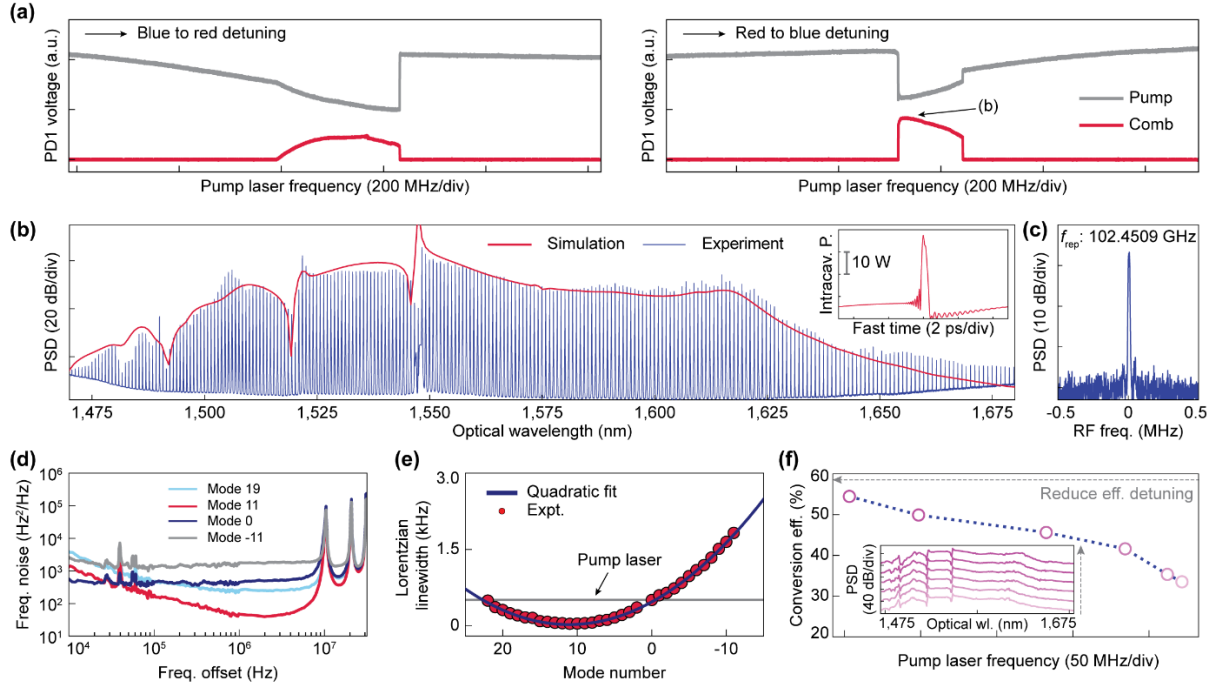

**Fig. 3 | Normal-dispersion Kerr microcomb from the 102.4 GHz-FSR device. (a)** Pump and comb powers represented by gray and red curves, respectively, as the pump laser is swept from blue to red (left panel) and red to blue (right panel). The curves are vertically offset for clarity. **(b)** Measured (blue) and simulated (red) spectrum corresponding to the state labeled "(b)" in the right panel of Fig. 3(a). The inset shows the simulated temporal waveform. **(c)** Repetition-rate beatnote electro-optically down converted to the microwave domain. The measurement span is 1 MHz, and the resolution bandwidth is 250 Hz. **(d)** Frequency noise power spectral density of selected lines of the comb in Fig. 3(b). **(e)** Lorentzian (intrinsic) linewidths of 33 comb lines (red) about the pump mode (designated mode 0). The Lorentzian linewidth of mode 0 is marked by the horizontal gray line. The quadratic fit (blue) shows the mode with minimum Lorentzian linewidth is blue-shifted from the pump. Mode 1 data is neglected, since this mode had insufficient signal-to-noise ratio. **(f)** Measured pump-to-comb conversion efficiency, increasing from ~33% to ~54%. Inset shows spectral envelopes of the corresponding spectra. Gray arrows indicate the direction of smaller effective detuning achieved via decreasing the pump laser frequency.

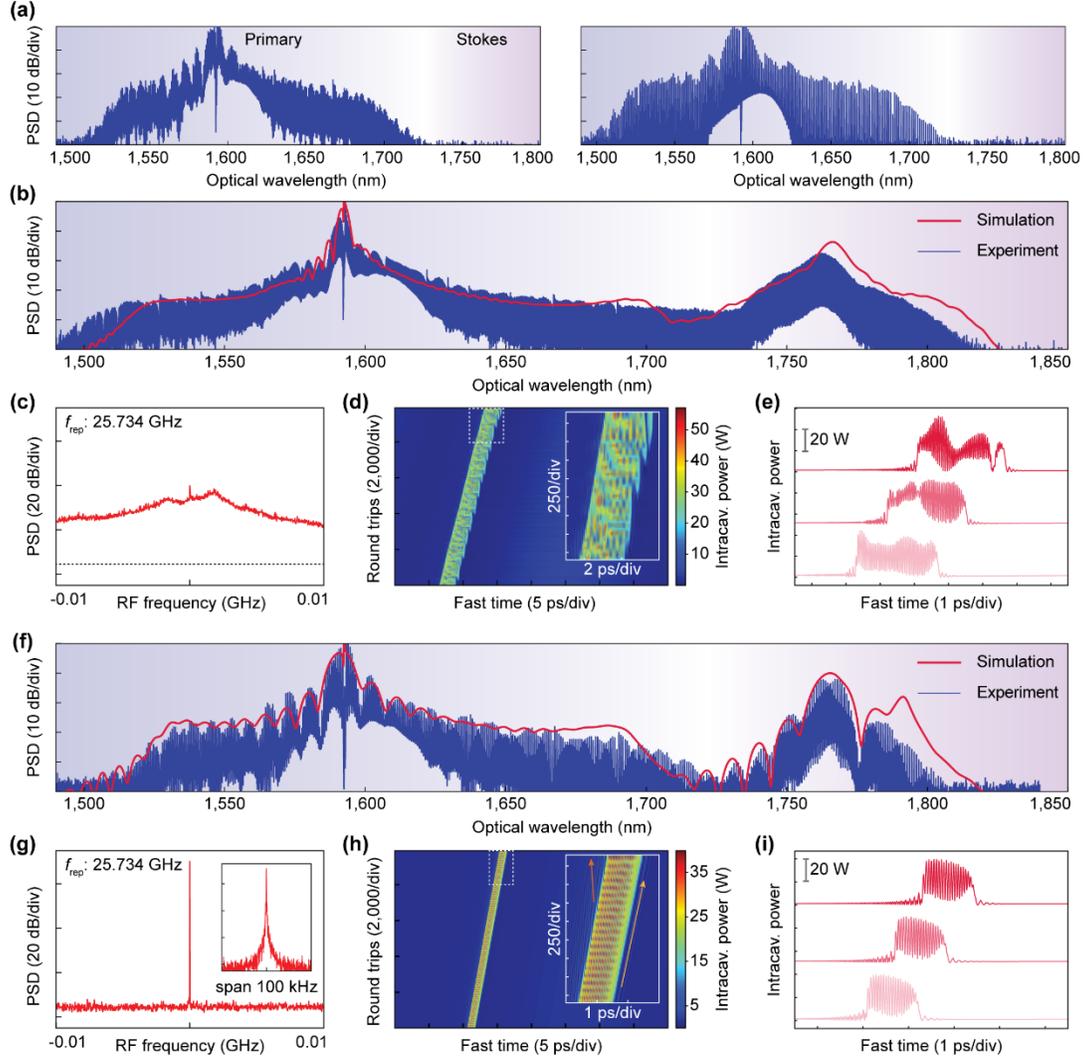

**Fig. 4 | Normal-dispersion Kerr and Raman Stokes microcombs from the 25.7 GHz-FSR device. (a)** Normal-dispersion Kerr microcombs with varied effective repetition-rates (left panel, $f_{rep} \sim 1 \cdot f_{FSR}$; right panel, $f_{rep} \sim 5 \cdot f_{FSR}$). **(b)** Measured (blue) and simulated (red) spectrum of a chaotic state. **(c)** Repetition-rate beatnote of the comb in (b). The measurement span is 20 MHz, and the resolution bandwidth (RBW) is 1 kHz. **(d)** Simulated round-trip-to-round-trip evolution of the chaotic state for 10,000 round trips. The inset shows a zoomed-in view over the last 2,000 round trips, exhibiting no discernable pattern or repeatability. **(e)** Simulated temporal waveforms at selected round trips, showing drastic changes between round trips. **(f)** Measured (blue) and simulated (red) spectrum of a temporally shorter stable state, compared to (b). **(g)** Repetition-rate beatnote of the comb in (f), with same measurement settings as (c). The inset shows a zoomed-in view with 100 kHz span and 100 Hz RBW. **(h)** Simulated round-trip-to-round-trip evolution of the stable state for 10,000 round trips. The inset shows a zoomed-in view over the last 2,000 round trips, and the color gradient clearly shows temporal walk-off between the flat-top pulse and its oscillatory feature, induced by phase-velocity mismatch between the pump and its Stokes-shifted wavelengths. Orange arrows illustrate the walk-off and serve as a guide to the eye. **(i)** Simulated temporal waveforms at selected round trips, showing good stability. In (e) and (i), the waveforms are vertically offset for clarity. In (a), (b), and (f), the blue and red backgrounds indicate the approximate spectral regions of the primary Kerr and Raman Stokes microcombs, respectively.